\documentclass[twocolumn,letterpaper]{IEEEconf}
\usepackage[left=0.75in,top=0.72in,right=0.75in,bottom=0.78in]{geometry}

\usepackage{amsmath}
\usepackage{amssymb}
\usepackage{graphicx}
\usepackage{subfigure}
\usepackage{verbatim}
\usepackage[thinlines,thiklines]{easybmat}
\usepackage{latexsym}
\usepackage[dvipsnames]{xcolor}
\usepackage{cite}
\usepackage{bm}
\usepackage{stmaryrd}
\usepackage{multirow}
\usepackage{textcomp}
\usepackage{bookmark}
\usepackage{booktabs}
\usepackage{balance}
\usepackage{stmaryrd}
\usepackage{mathtools}
\usepackage{stfloats}
\usepackage{epstopdf}

\def\diag{\mathop{\rm diag}\nolimits}

\newcommand{\bs}[1]{\ensuremath{{\boldsymbol{#1}}}}

\usepackage{algorithm}
\usepackage{algorithmic}

\def\atan2{\mathrm{atan2}}

\begin{document}

\title{\bf \Large Machine Learning-based Online Stability Lobe Diagram Estimation and \\ Chatter Suppression Control in Milling Process~\thanks{The work was supported in part by the US National Science Foundation (NSF) under award ECCS-2328260.}}

\author{Yi Huang\thanks{Y. Huang, W. Liu, J. Yi, and Y. Guo are with the Department of Mechanical and Aerospace Engineering, Rutgers University, Piscataway, NJ 08854 USA (e-mail: \{easton.hy, wenyi. liu, jgyi, yuebin.guo\}@rutgers.edu).}, Feng Han\thanks{F. Han is with the Department of Mechanical Engineering, New York Institute of Technology, Old Westbury, NY 11568 USA (e-mail: fhan03@nyit.edu).}, Wenyi Liu, Jingang Yi, and Yuebin Guo}

\maketitle

\begin{abstract}
Chatter is a self-excited vibration in milling that degrades surface quality and accelerates tool wear. This paper presents an adaptive process controller that suppresses chatter by leveraging machine learning-based online estimation of the Stability Lobe Diagram (SLD) and surface roughness in the process. Stability analysis is conducted using the semi-discretization method for milling dynamics modeled by delay differential equations. An integrated machine learning framework estimates the SLD from sensor data and predicts surface roughness for chatter detection in real time. These estimates are integrated into an optimal controller that adaptively adjusts spindle speed to maintain process stability and improve surface finish. Simulations and experiments are performed to demonstrate the superior performance compared to the existing approaches.
\end{abstract}

\section{Introduction}

Chatter is a self-excited vibration that occurs during the milling process, adversely affecting machining efficiency and product quality~\cite{24_CIRP_chatter, 25_JMP_Ding}. To address this challenge, extensive research work have been conducted on chatter prediction and suppression. A common approach involves transforming time-series data into the frequency domain to capture spectral characteristics~\cite{21_CIRP_Freq, 22_MSSP_freq}. Machine learning models have recently been used to capture the nonlinear dynamics of machining~\cite{24_JDSMC_feng, 21_RAL_feng, 25_TASE_siyu}, enabling chatter prediction through indicators like surface roughness~\cite{24_PR_liwen_R, 23_ML_liwen_R, 25_JMSE_liwen}. To address the limited physical interpretability of these models, physics-informed machine learning has been introduced, improving prediction accuracy with reduced training data~\cite{25_ACC_Yi, 19_CIRP_PINNs, 25_JMP_PINNs}.

The milling process is commonly modeled using delay differential equation (DDE). Chatter arises from the regenerative effect caused by the delayed term, and is therefore analyzed through the stability of the DDE-based dynamics model~\cite{22_MSSP_ding}. Since DDE depends on both current and past states, they are infinite-dimensional. To address this, discretization is employed to transform them into finite-dimensional systems for numerical stability analysis. Full-discretization approaches yield highly accurate stability predictions by retaining the complete delay history~\cite{10_IJMTM_Fulldisc}, but are computationally expensive. In contrast, semi-discretization methods (SDM) offer a compromise approach by using a limited set of past states to balance accuracy and efficiency~\cite{11_Book_semi, 21_IJAMT_SDM}. Stability lobe diagram (SLD), derived from DDE models, delineates stable and unstable cutting regions based on spindle speed and depth of cut. Efficient and accurate SLD estimation is essential for chatter avoidance.

Approaches to chatter suppression are generally categorized as active or passive control, both aiming to shift the process into stable regions on the SLD. Active chatter suppression leverages actuators with real-time feedback control to expand the stable region, enabling previously unstable parameters~\cite{24_IJAMT_electromagnetic, 19_MSSP_piezoelectric, 15_IJMTM_MPC}. However, these methods rely heavily on accurate modeling and precise controller tuning, while additional actuators increase system complexity and uncertainty~\cite{25_MSSP_ActiveSaturation}. Passive control, in contrast, is simpler and more robust. Adaptive spindle speed tuning, a widely used passive strategy, aims to prevent chatter by optimizing milling spindle speed based on an offline-identified SLD model~\cite{17_IECON_Fujimoto, 24_CIRP_SSV, 20_JMP_BayeLearning}. Cost-function-based passive controllers have also demonstrated effectiveness in chatter suppression~\cite{17_IECON_Fujimoto, 19_IJMTM_Cost}.

Existing passive chatter control methods typically rely on a fixed, offline-identified SLD and milling dynamics model, with spindle speed preset and held constant during machining~\cite{17_AIM_Fujimoto, 17_IECON_Fujimoto}. However, variations in assembly, spindle type, workpiece material, and setup can shift the cutting dynamics and alter the SLD. As a result, fixed SLD-based speed selection may become suboptimal or unstable, leading to chatter and poor surface finish. In addition, online estimation of system parameters for analytically SLD updating in real-time remains challenging, as these parameters typically require dedicated experiments for identification and are unmeasurable during milling.

This paper proposes a machine learning-based method for online SLD estimation and passive chatter control. Since the system parameters are unmeasurable during milling, a novel loss function is formulated by embedding the dynamics model to compute measurable cutting forces. This model-driven approach enables indirect learning of parameter variations and supports real-time adaptation of spindle speed to dynamic process conditions. Moreover, even within the stable region of the SLD, spindle speed significantly affects the resulting surface roughness. To address this, a cost-function-based controller is therefore developed to optimize spindle speed for both guaranteed stability and surface quality, thereby enhancing chatter suppression effectiveness. Both simulations and experiments validate the effectiveness of the proposed scheme, with comparative analysis demonstrating its superiority over existing controllers for adaptive spindle speed selection. The main contribution of this work lies in the new machine learning-enabled chatter suppression control design that is adapatively adjusted to process conditions in real time.

The remainder of the paper is organized as follows. We present the problem statement in Section~\ref{sec_prostate}. Section~\ref{sec_SDM} presents the SDM for milling dynamics and analysis with DDEs. The adaptive process controller are described in Section~\ref{sec_controller_design}. The experiments and results are presented in Section~\ref{sec_exp}. Concluding remarks are summarized in Section~\ref{sec_Con}.

\section{Problem Statement and Approach Overview}
\label{sec_prostate}

\subsection{Problem Statement}
\label{subsec_problem_state}

\begin{figure}[h!]
  \centering
  \includegraphics[width=0.98\linewidth]{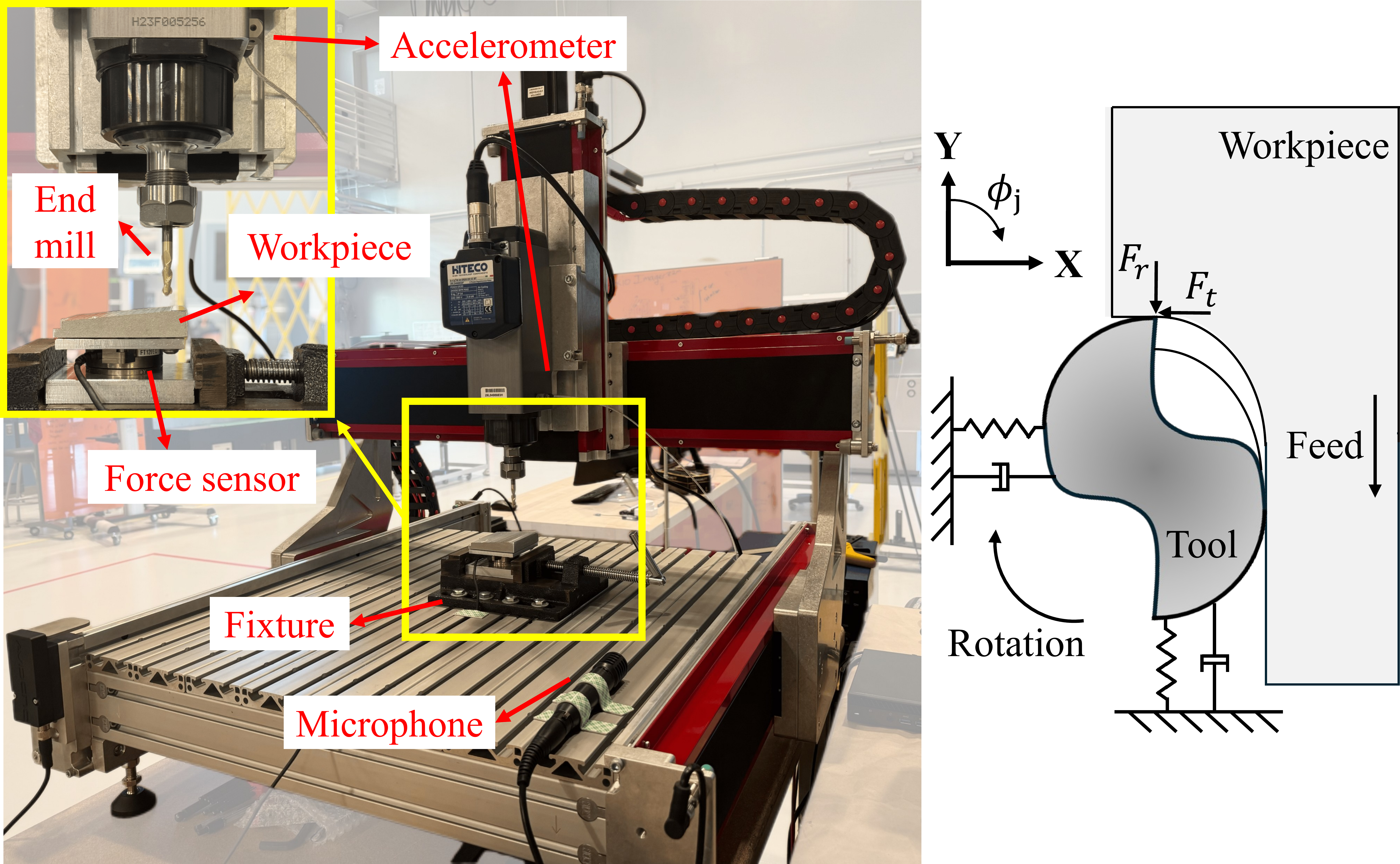}
  \caption{Experimental setup and schematics of the milling process. Left: Milling machine used in experiments. Right: Modeling schematic.}
  \label{fig_exp_setup}
  \vspace{-0mm}
\end{figure}

Fig.~\ref{fig_exp_setup} shows the physical setup for the milling process and also illustrates a simplified two-degree-of-freedom (2-DoF) dynamic model in the $XY$-plane. Since the workpiece is rigidly clamped, the chatter analysis focuses solely on the spindle dynamics. Denoting the 2-DoF displacement of the spindle as $\bm q(t)=[q_x(t)\; q_y(t)]^T \in \mathbb{R}^2$, the milling dynamics are formulated as~\cite{25_ACC_Yi}:
\begin{equation}
    \label{eq_dynamic_model}
    \bm M \ddot{\bm q}(t) + \bm C\dot{\bm q}(t) + \bm K \bm q(t) = \bm F(t),
\end{equation}
where $\bm M=\diag(M_x, M_y)$, $\bm C=\diag(2\zeta\omega_n M_x, 2\zeta\omega_n M_y)$ and $\bm K=\diag(\omega_n^2 M_x, \omega_n^2 M_y)$ are the mass, damping, and stiffness matrices, respectively, and $\omega_n$ denotes as natural frequency of the system. The cutting force $\bm F(t)=\bm F_s(t)+\bm F_d(t)$ consists of a static component $\bm F_s(t)$ and a dynamic component $\bm F_d(t)$.

The dynamic force is modeled by $\bm F_d(t)  =a_p\bm H_d[\bm q(t)-\bm q(t-\tau)]$ and is a regenerative force component due to the delay difference between $t$ and $t-\tau$, where $\bm H_d(t)$ represents the time-varying cutting force coefficient matrix. Given the number of teeth $N$ and the spindle speed $\omega_{sp}$, the regenerative delay is defined as $\tau=\frac {60}{N\omega_{sp}}$, which is a primary contributor to chatter during the milling process. Consequently,~\eqref{eq_dynamic_model} is reduced to a perturbation system that excludes the static force $\bm F_s$. 

Incorporating the expression for $\bm F_d$ and the active control input $\bm u$, the resulting perturbation dynamics are described as:
\begin{align}
    \bm M \ddot{\bm q}(t)+ \bm C\dot{\bm q}(t) +\bm K \bm q(t) = & a_p\bm H_d(t)[\bm q(t)-\bm q(t-\tau)] \nonumber \\ 
    & + \bm u(t),
    \label{eq_perturbation_model}
\end{align}
where $\bm u(t)$ denotes the control input from actuators and $a_p$ is the depth of cut. The matrix $\bm H_d(t)$ is given by
\begin{equation}
\label{eq_H_d}
    \bm{H}_d(t) =\frac{1}{2} \sum_{j=0}^{N-1}g(\phi_j(t))
    \begin{bmatrix}
    \eta_{xx}(\phi_j(t)) & \eta_{xy}(\phi_j(t)) \\
    \eta_{yx}(\phi_j(t)) & \eta_{yy}(\phi_j(t))
    \end{bmatrix},
\end{equation}
where 
\begin{align*}
  & \eta_{xx}(\phi_j(t))=K_t \sin(2\phi_j(t))+K_r[1-\cos(2\phi_j(t))], \\
  & \eta_{xy}(\phi_j(t))=K_t [1+\cos(2\phi_j(t))]+K_r\sin(2\phi_j(t)),  \\
  & \eta_{yx}(\phi_j(t))=-K_t [1-\cos(2\phi_j(t))]+K_r\sin(2\phi_j(t)), \\
  & \eta_{xx}(\phi_j(t))=-K_t \sin(2\phi_j(t))+K_r[1+\cos(2\phi_j(t))],  
\end{align*}
and $\phi_j(t) = ( \frac{2\pi \omega_{sp}}{60})t + (j - 1) \frac{2\pi}{N}$ is the $j$-th tooth's angle. Tooth engagement spans from the entry angle $\phi_{\text{in}} = \arccos\left(1 - \frac{2a_e}{D}\right)$ to exit angles $\phi_{\text{out}}= \pi$, where the $D$ is the diameter of end mill and $a_e$ is the radial of depth of cut. Term $g(\phi_j(t))$ in~\eqref{eq_H_d} is a switching function that equals 1 if the $j$-th tooth lies within the cutting interval $(\phi_{in}, \phi_{out})$, and 0 otherwise; $K_t$ and $K_r$ denote the coefficients of tangential $F_t$ and radial $F_r$ cutting force, respectively.

During machining, the regenerative force would cause chattering and thereby reduce the surface finishing quality. Active control design is needed to mitigate the regenerative force and thus suppress chattering. In~\eqref{eq_perturbation_model}, the system coefficients capture the dynamic properties of the tool and the workpiece. Accurate measurement of $\bm M$, $\bm C$, and $\bm K$ requires complex experimentation and system identification~\cite{19_MSSP_piezoelectric, 24_IJAMT_electromagnetic}. In addition, the tool installation, workpiece surface roughness, and other in-mission cutting conditions would change the parameters. The one-time identified parameters in a specific setup, if available, would vary during the machining process. In this work, we recognize that the system parameters $\bm M$, $\bm C$, and $\bm K$ are not available. Therefore, active control design to suppress the chattering and achieve fine and high-quality surface finishing becomes challenging.

{\em Problem Statement}: The objective of this work is to design a real-time controller capable of suppressing chatter without requiring identification of system coefficients, thereby enhancing milling efficiency while maintaining high surface finishing quality with low roughness.

\subsection{Approach Overview}
\label{subsec_approach_overview}

\begin{figure}[h!]
	\centering
	\vspace{2mm}
	\includegraphics[width=0.98\linewidth]{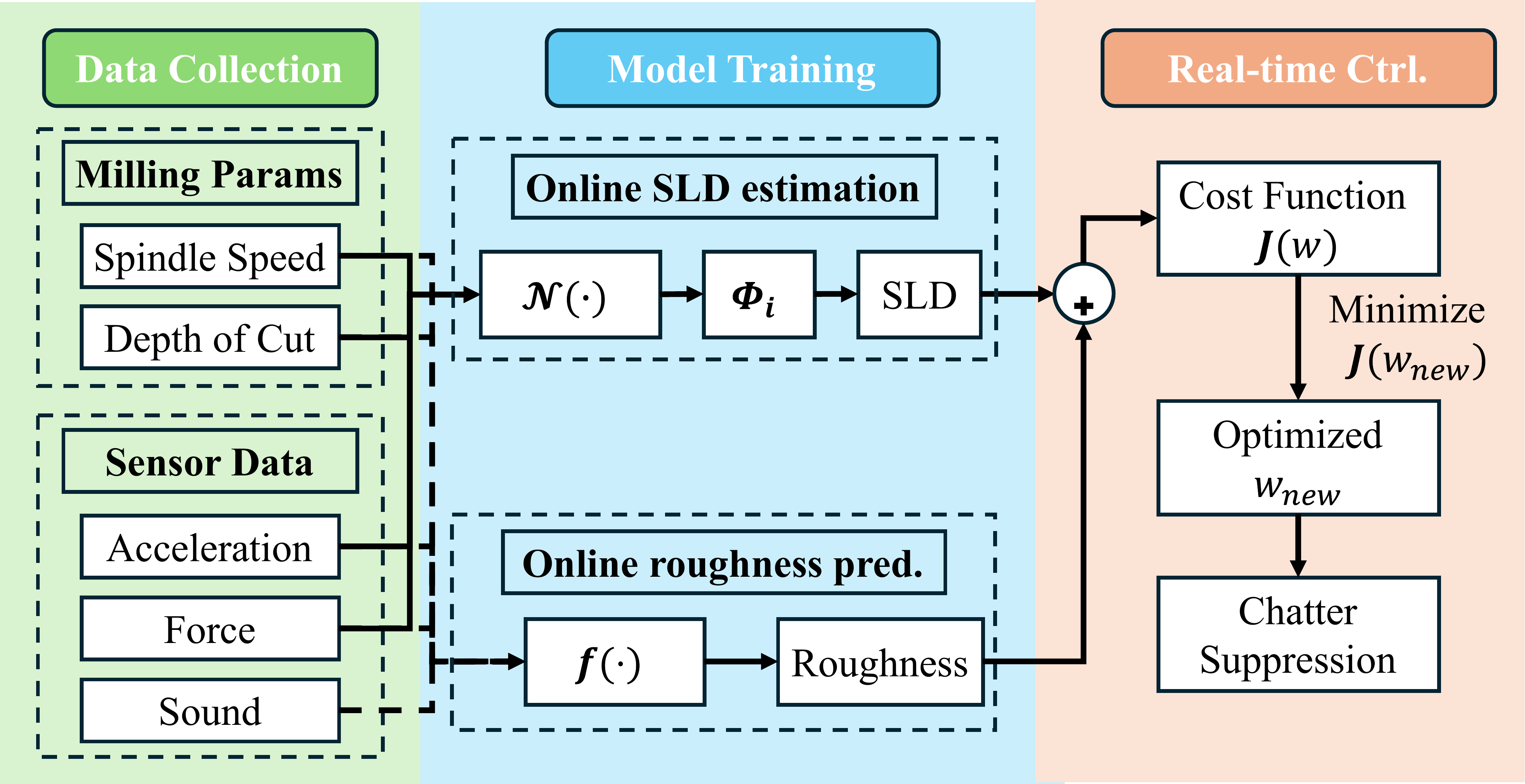}
	\caption{Overview of the proposed chatter suppression control strategy.}
	\label{fig_overview}
	\vspace{-1mm}
\end{figure}

The proposed approach integrates a machine learning-based online estimations with an adaptive process controller design for real-time chatter suppression. Fig.~\ref{fig_overview} illustrates the overview of the proposed chatter suppression control strategy. The real-time controller is structured into three key components: (1) Online SLD estimation, which identifies the stable cutting region based on real-time system behavior, enabling continuous spindle speed adaptation; (2) Online roughness prediction supports real-time chatter detection and serves as a reference for evaluating machining quality; (3) A cost function design combines stability and surface roughness metrics to evaluate cutting performance and guide the optimization of spindle speed for effective chatter suppression. In the next two sectcions, we will describe the online SLD estimation and then the chatter suppression design. 

\section{Stability Analysis and Learning-based Online Estimation}
\label{sec_SDM}

In this section, stability analysis of the milling process is conducted, and a machine learning-based method for online SLD estimation is proposed based on the analysis.

\subsection{Stability Analysis Based on Semi-Discretization}

To facilitate effective controller design, the regenerative chatter dynamics in~\eqref{eq_perturbation_model} is reformulated into a linear time-varying (LTV) system, expressed in the following state space 
\begin{equation}
\label{eq_state_represent_continous}
  \dot{\bm{x}}(t) = \bm{A}(t)\bm{x}(t) + \bm{B}(t)\bm{x}(t - \tau)+ \bm{D}(t) \bm u(t),
\end{equation}
where the state vector is $\bm x(t) = [\bm q(t)^T \quad \dot{\bm q}(t)^T]^T \in \mathbb{R}^{4}$, control input vector $\bm u(t) \in \mathbb{R}^{2}$, and matrices $\bm A(t) \in \mathbb{R}^{4 \times 4}$, $\bm B(t) \in \mathbb{R}^{4 \times 4}$, and $\bm D(t) \in \mathbb{R}^{4 \times 2}$ are given by
\begin{gather*}
  \bm A(t) = \begin{bmatrix}
    \bm{O}_{2,2} & \bm{I}_{2} \\
    -\bm{M}^{-1}(\bm{K} + a_p \bm{H}_d(t)) & -\bm{M}^{-1} \bm{C}
    \end{bmatrix}, \\
  \bm B(t) = \begin{bmatrix}
    \bm{O}_{2,2} & \bm{O}_{2,2} \\
    a_p \bm{M}^{-1} \bm{H}_d(t) & \bm{O}_{2,2}
    \end{bmatrix},
  \bm D(t) = \begin{bmatrix}
    \bm{O}_{2,2} \\
    a_p \bm{M}^{-1} \bm{H}_d(t)
    \end{bmatrix},
\end{gather*}
$\bm O_{k,n} \in \mathbb{R}^{k \times n}$ is the zero matrix and $\bm I_{n} \in \mathbb{R}^{n \times n}$ is the identity matrix. To analyze the stability of the DDE presented in~\eqref{eq_state_represent_continous}, the SDM is employed~\cite{11_Book_semi}. This method provides a numerically stable and accurate approach for creating SLD.

{\em Local Stability via SDM}: Given sampling period $\epsilon$, the regenerative delay is approximated as $\tau= m \epsilon$, $ m \in \mathbb{N}$. For $i$-th interval, $t _i \in [i \epsilon, (i + 1) \epsilon]$,~\eqref{eq_state_represent_continous} becomes
\begin{equation}\label{eq_state_appro}
    \dot{\bm{x}}(t_i) = \bm{A}_i \bm{x}(t_i) + \bm{B}_i \bm{x}_{\tau,i}+ \bm{D}_i \bm u(t_i),
\end{equation}
where $\bm{A}_i = \frac{1}{\epsilon} \smallint_{i\epsilon}^{(i+1)\epsilon} \bm{A}(t)dt$, $\bm{B}_i = \frac{1}{\epsilon} \smallint_{i\epsilon}^{(i+1)\epsilon} \bm{B}(t)dt$, and $\bm{D}_i = \frac{1}{\epsilon} \smallint_{i\epsilon}^{(i+1)\epsilon} \bm{D}(t)dt$. The time delay term is approximated as $\bm x(t_i-\tau) \approx \bm x (i\epsilon+\frac{\epsilon}{2}-\tau) = \frac{1}{2}(\bm x_{i-m} + \bm x_{i-m+1})$. Introducing the notation $\bm{x}(t_i) = \bm{x}_i$ and $\bm{u}(t_i) = \bm{u}_i$, and assuming that $\bm{u}(t)$ is constant in $t_i \in [i \epsilon, (i + 1) \epsilon]$, integration of~\eqref{eq_state_appro} over the interval and evaluation at $t=t_{i+1}$ yields
\begin{equation}\label{eq_state_discr}
    \bm{x}_{i+1} = \bm{P}_i \bm{x}_i + \frac{1}{2} \bm{R}_i (\bm x_{i-m} + \bm x_{i-m+1}) + \bm{Q}_i \bm{u}_i,
\end{equation}
where $\bm{P}_i = \exp(\bm{A}_{i} \epsilon), \bm{R}_i = \left( \exp(\bm{A}_{i} \epsilon) - \bm{I}_{2} \right) \bm{A}_{i}^{-1} \bm{B}_{i}\in \mathbb{R}^{4 \times 4}$, and  $\bm{Q}_i = \left( \exp(\bm{A}_{i} \epsilon) - \bm{I}_{2} \right) \bm{A}_{i}^{-1} \bm{D}_i \in \mathbb{R}^{4 \times 2}$.

Based on~\eqref{eq_state_discr}, the state vector $\bm x_i$ is augmented to incorporate historical state information, yielding the extended state $\bm y_i= \operatorname{col}(\bm{q}^T_i \ \dot{\bm{q}}^T_i \ \bm{q}^T_{i-1} \ \dots \ \bm{q}^T_{i-m}) \in \mathbb{R}^{2(m+2)}$. Therefore,~\eqref{eq_state_discr} is rewritten with the augmented state as
\begin{subequations}\label{eq_state_augmented}
    \begin{align}
      \bm y_{i+1}&= \bm E_i \bm y_{i} + \bm G_i \bm u_i, \label{eq_state_augmented} \\
        \bm{E}_i &= \left[
        \begin{array}{c|c}
        \bm{P}_i & \bm{T}_i \\
        \hline
        \bm{S} & \bm{O}_{2,2m} \\
        \hline
        \bm{O}_{2m-2,4} & \bm{I}_{2m}
        \end{array}
        \right], \;
        \bm{G}_i = \begin{bmatrix}
            \bm{D}_i \\
            \bm{O}_{2m, 2}
            \end{bmatrix},\label{eq_state_augment_E_G}
    \end{align}
\end{subequations}
where matrices $\bm E_i \in \mathbb{R}^{2(m+2) \times 2(m+2)}$ and $\bm G_i \in \mathbb{R}^{2(m+2) \times 2}$, with $\bm{S} = [ \bm{I}_{2} \ \bm{O}_{2,2}]$. Consequently, matrices $\bm{T}_i = \left[ \bm{O}_{4,2k-4} \ \tfrac{1}{2} \bm{R}_i^{(4,2)} \ \tfrac{1}{2} \bm{R}_i^{(4,2)} \right]$, where $\bm R_i^{(k,n)}$ denotes the submatrix of $\bm R_i$ consisting of first $k$ rows and $n$ columns.

{\em Global Stability and Chatter Prediction}:~\eqref{eq_state_augmented} discretizes the single $\tau$-period system into $m$ steps, yielding a finite-dimensional approximation. In contrast, the DDE described in~\eqref{eq_perturbation_model} is inherently infinite-dimensional due to its dependence on past states over a continuous time interval. To bridge this gap and facilitate global analysis over periods, a transition matrix $\bm \Phi$ is introduced to evaluate system evolution across the entire period,
\begin{subequations}
\label{eq_state_global}
    \begin{align}
        & \bm y_{i+m}= \bm \Phi_i \bm y_{i} + \bm \Gamma_i \bm u_{i},\label{eq_state_global_main}\\
        & \bm{\Phi}_i = \bm{E}_{i+m-1} \bm{E}_{i+m-2} \cdots \bm{E}_i,\label{eq_state_global_phi}\\
        & \bm{\Gamma}_i = \sum\nolimits_{j=0}^{m-1} \left( \prod\nolimits_{l=j+1}^{m-1} \bm{E}_{i+l} \right) \bm{G}_{i+j},\label{eq_state_global_gamma}
    \end{align}
\end{subequations}
where $\bm{\Phi}_i \in \mathbb{R}^{2(m+2) \times 2(m+2)}$ and $\bm{\Gamma}_i \in \mathbb{R}^{2(m+2) \times 2}$.

Within the semi-discretization framework, the global transition matrix $\bm \Phi_i$ maps the augmented state across one delay period $\tau$. Consequently, the onset of chatter is determined by the spectral properties of $\bm \Phi_i$: if its maximum singular value $\lambda_{\max}(\bm \Phi_i)>1$, the system becomes unstable, leading to regenerative chatter. Conversely, if all singular values remain $\lambda_{\max}(\bm \Phi_i)<1$, the system is asymptotically stable and chatter is suppressed. To analyze the stability of a specific dynamic system with all parameters identified, matrix $\bm \Phi_i$ is expressed as a function of spindle speed and depth of cut, denoted as $\bm \Phi_i(\omega_{sp}, a_p)$. Chatter suppression is achieved by selecting spindle speed and depth of cut combinations that lie within the stable region.

While estimating matrix $\bm{\Phi}_i$ and analyzing its singular values for stability analysis remains challenging, particularly when model parameters are unavailable. SLD provides a graphical representation of the maximum singular value of $\bm{\Phi}_i$. Meanwhile, considering the chip brings external force input $\bm u$, $\bm{\Gamma}_i$ is also closely related to chatter. However, given the assumption of constant chip thickness, it is sufficient to use $\bm{\Gamma}_1$, which reflects only the current input influence without accounting for past cutting states~\cite{17_IECON_Fujimoto}. Therefore, we propose real-time SLD estimation to guide spindle speed modulation in the control design.

\subsection{Learning-based Online Estimation of SLD}

Based on the stability analysis, the SLD is characterized by the maximum eigenvalue of $\bm \Phi_i = \bm \Phi_i(\omega_{sp}, a_p)$ in~\eqref{eq_state_global}. The state transition matrix $\bm \Phi_i$ depends on both fixed and uncertain system parameters. Parameters such as the number of teeth, mass, end mill diameter, and radial depth of cut are readily measurable and remain constant during milling. In contrast, the damping ratio $\zeta$, natural frequency $\omega_n$, and cutting force coefficients $K_t$ and $K_r$ are difficult to measure and may vary due to thermal effects, vibration, and other uncertainties. Therefore, these uncertain parameters are treated as variables in the formulation of $\bm \Phi_i$, and the online stability condition is further defined with the maximum eigenvalue of the $\bm \Phi_i$, yielding stability if 
\begin{equation}
\label{eq_SLD_Phi}
\lambda_{\max}(\bm \Phi_i(\underbrace{\xi, \omega_n, K_t, K_r}_{\text{System Param.}}; \underbrace{a_p, \omega_{sp}}_{\text{Variable}}))<1.
\end{equation}

\begin{figure}[h!]
  \centering
  \includegraphics[width=0.98\linewidth]{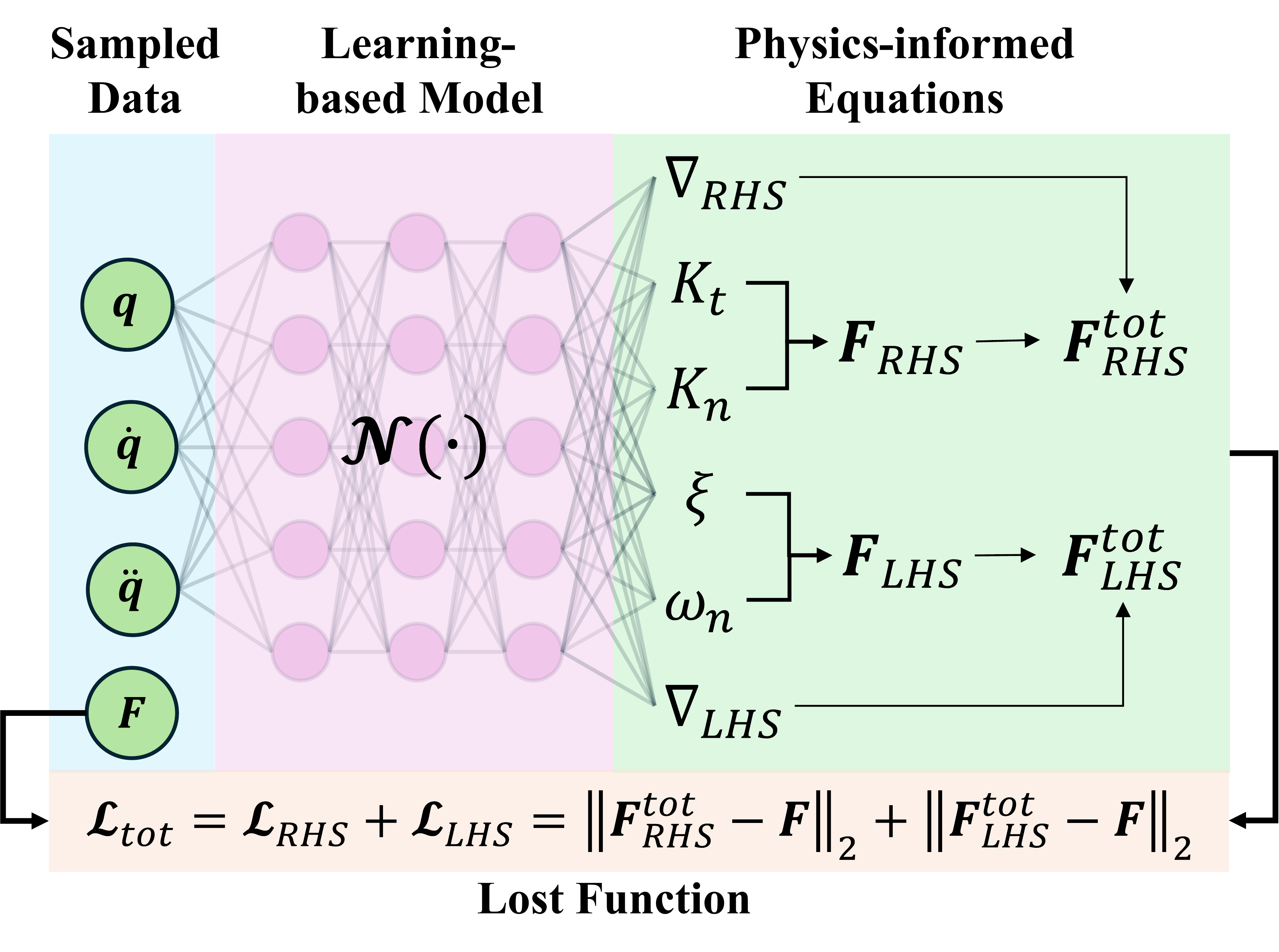}
  \vspace{-3mm}
  \caption{Flowchart of the machine learning-based approach for real-time SLD estimations.}
  \label{fig_SLD_model}
  \vspace{-0mm}
\end{figure}

In order to estimate the system parameters during milling, which evolve dynamically with the process, a learning-based model is developed that utilizes online sensor data to capture and predict these changes in real time. Fig.~\ref{fig_SLD_model} illustrates the overview of the machine learning-based SLD estimation. Instead of designing the loss function based on unmeasurable system parameters during milling, a learning-based model is trained by embedding~\eqref{eq_perturbation_model} with predicted system parameters to compute the force on both sides of the equation and compare with the measured force. The machine learning-based model is represented as a function $\mathcal{N}(\cdot)$, implemented through multiple layers of a neural network. The input to $\mathcal{N}(\cdot)$ is the motion state of the end mill, including the acceleration $\ddot{\bm q}(t)$, velocity $\dot{\bm q}(t)$, and displacement $\bm q(t)$, and the outputs consist of both the system parameters and correction terms.

Directly estimating the system parameters is challenging because these quantities are difficult to measure and record as labeled data for training. To address this, the loss function is formulated based on the perturbation model in~\eqref{eq_perturbation_model}. On the left-hand side of this equation, the system parameters of interest include the damping ratio $\zeta (t)$ and the natural frequency $\omega_{n}(t)$, which define matrices $\bm C(\zeta(t), \omega_{n}(t))$ and $\bm K(\omega_{n}(t))$. On the right-hand side, matrix $\bm H_d$ is expressed in terms of the tangential and radial cutting force coefficients, $K_t(t)$ and $K_r(t)$ as $\bm H_d(K_t(t), K_r(t))$. The forces on both sides of~\eqref{eq_perturbation_model}, denoted $\bm F_\mathrm{LHS}$ and $\bm F_\mathrm{RHS}$, are computed as
\begin{subequations}
\label{eq_model_force}
\begin{align*}
  \bm F_\mathrm{LHS}&= \bm M \ddot{\bm q}(t) + \bm C\left(\zeta(t), \omega_{n}(t)\right)\dot{\bm q}(t) + \bm K\left(\omega_{n}(t)\right) \bm q(t), \\
  \bm F_\mathrm{RHS} &= a_p\bm H_d \left( K_t(t), K_r(t) \right)[\bm q(t)-\bm q(t-\tau)].
\end{align*}
\end{subequations}

To account for uncertainties, correction terms, denoted by $\nabla$, are added to both sides, yielding the total forces $\bm F_\mathrm{LHS}^\mathrm{tot}$ and $\bm F_\mathrm{RHS}^\mathrm{tot}$, which are compared with the measured force $\bm F$. The total loss $\mathcal{L}_\mathrm{tot}$ is defined as the sum of the individual loss functions, i.e., $\mathcal{L}_\mathrm{tot}=\mathcal{L}_\mathrm{LHS}+\mathcal{L}_\mathrm{RHS}$, and is expressed as
\begin{equation}\label{eq_SLD_loss}
    \mathcal{L}_\mathrm{tot}
    = \| \underbrace{\bm F_\mathrm{LHS} + \nabla_{LHS}}_{\bm F_\mathrm{LHS}^\mathrm{tot}} - \bm{F} \|_2^2
    + \| \underbrace{\bm F_\mathrm{RHS} + \nabla_\mathrm{RHS}}_{\bm F_\mathrm{RHS}^\mathrm{tot}} - \bm{F} \|_2.
\end{equation}

By training the learning-based model $\mathcal{N}\left( \bm q(t), \dot{\bm q}(t), \ddot{\bm q}(t) \right)$ with sensor while indirectly incorporating physics-informed equations, the system parameters are predicted, and the stability is subsequently defined according to~\eqref{eq_SLD_Phi}.

\section{Real-time Adaptive Process Control}
\label{sec_controller_design}

The predicted SLD provides insight into the stability of the machining process, enabling adjustment of spindle speed and other machining parameters to ensure stable operation. However, even under stable conditions, the surface roughness of the workpiece may still vary. In light of this, we incorporate both stability and surface roughness into the control design framework, treating them as qualitative and quantitative criteria, respectively.

\subsection{Chatter Detection based on Roughness Prediction}

\begin{figure}[h!]
  \centering
  \includegraphics[width=0.98\linewidth]{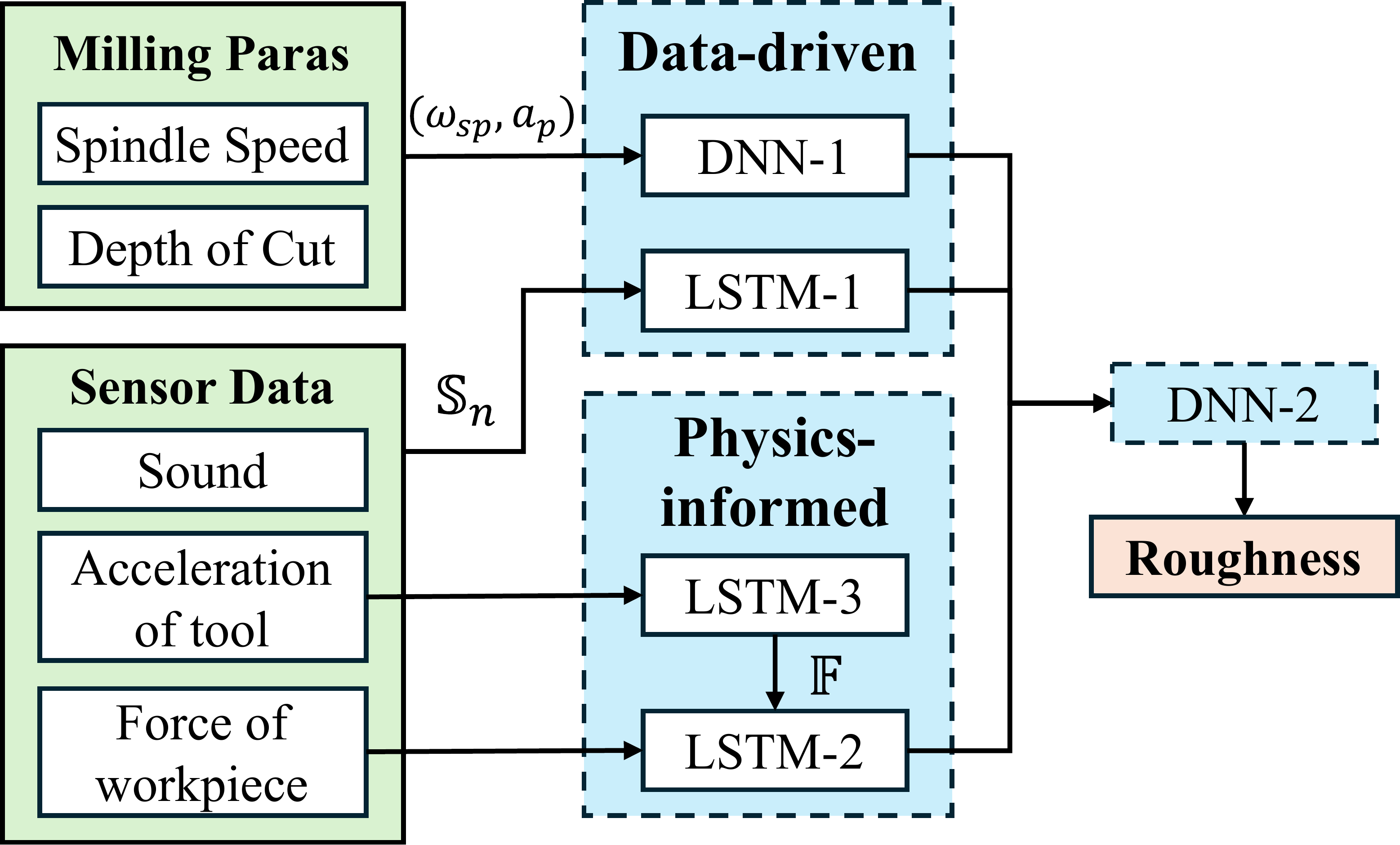}
  \caption{Flowchart of the machine learning-based roughness estimation.}
  \label{Fig_Roughnes}
  \vspace{-2mm}
\end{figure}

Real-time measurement of surface finishing roughness is typically unavailable due to the micrometer-level scale. Therefore, we employ a data-driven approach for roughness prediction, as detailed in~\cite{25_ACC_Yi}. Fig.~\ref{Fig_Roughnes} illustrates the architecture of the physics-informed neural network used for roughness prediction and chatter detection.

In milling process, the workpiece surface roughness, denoted by $r$, is a critical indicator for evaluating surface finishing quality. At the $(k+1)$th time step, $r_{k+1}$ is modeled as a function of several key variables:
\begin{equation}\label{Eq_F_Chatter}
r_{k+1}=f(d_{k}, \omega_{k}, p_k(d_{k}, \omega_{sp,k}), \mathbb{S}_n(k),\mathbb{F}(s_k)).
\end{equation}
Here, $d_k$ and $\omega_{sp,k})$ represents the milling depth and feed rate at time step $k$, respectively, and $p_k(d_k, \omega_k)$ denotes the probability distribution over milling parameter combinations that are estimated by a neural network. Term $\mathbb{S}_n(k) = \{\bs{s}_{k-n}, \bs{s}_{k-n+1}, \dots, \bs{s}_{k}\}$ is the sequence of sensor measurements from time steps $(k-n)$ to $k$, where $n \in \mathbb{N}$ and $n < k$. Each measurement $\bs{s}_k$ includes multi-modal signals such as acceleration, audio, and force data collected from the workpiece. The function $\mathbb{F}(s_k)$ represents the estimated milling force, derived from an offline-identified dynamics model using online tool acceleration measurements as input.

We consider a machine learning-based approach to estimate the function $f$ in~\eqref{Eq_F_Chatter}. As shown in Fig.~\ref{Fig_Roughnes}, the proposed surface roughness prediction framework comprises two deep neural networks (DNNs) and three long short-term memory (LSTM) models. The overall architecture integrates five neural networks to jointly predict surface roughness and chatter behavior: (1) {\em Parameter distribution estimation}: A DNN, denoted as {\tt DNN 1}, is trained to estimate the probability distribution $p(d,\omega_{sp})$ over various milling parameter combinations $(d,\omega_{sp})$; (2) {\em Feature extraction from raw signals}: An LSTM model, {\tt LSTM-1}, extracts temporal features from raw time-series measurement data; (3) {\em System dynamics modeling}: Using real-time acceleration measurements and dynamics model-based force estimation (with the obtained parameters in Fig.~\ref{fig_SLD_model}), the tool force is estimated. These forces, combined with workpiece interaction forces, are fed into {\tt LSTM-2} to extract dynamic behavior features of the entire milling system; (4) {\em Roughness prediction}: A second DNN, {\tt DNN 2}, takes the estimated probability $p(d, \omega_{sp})$ and the extracted features to predict the resulting surface roughness $r$.

\subsection{Adaptive Process Control}

\begin{figure}[h!]
	\centering
	\vspace{-1mm}
	\includegraphics[width=1\linewidth]{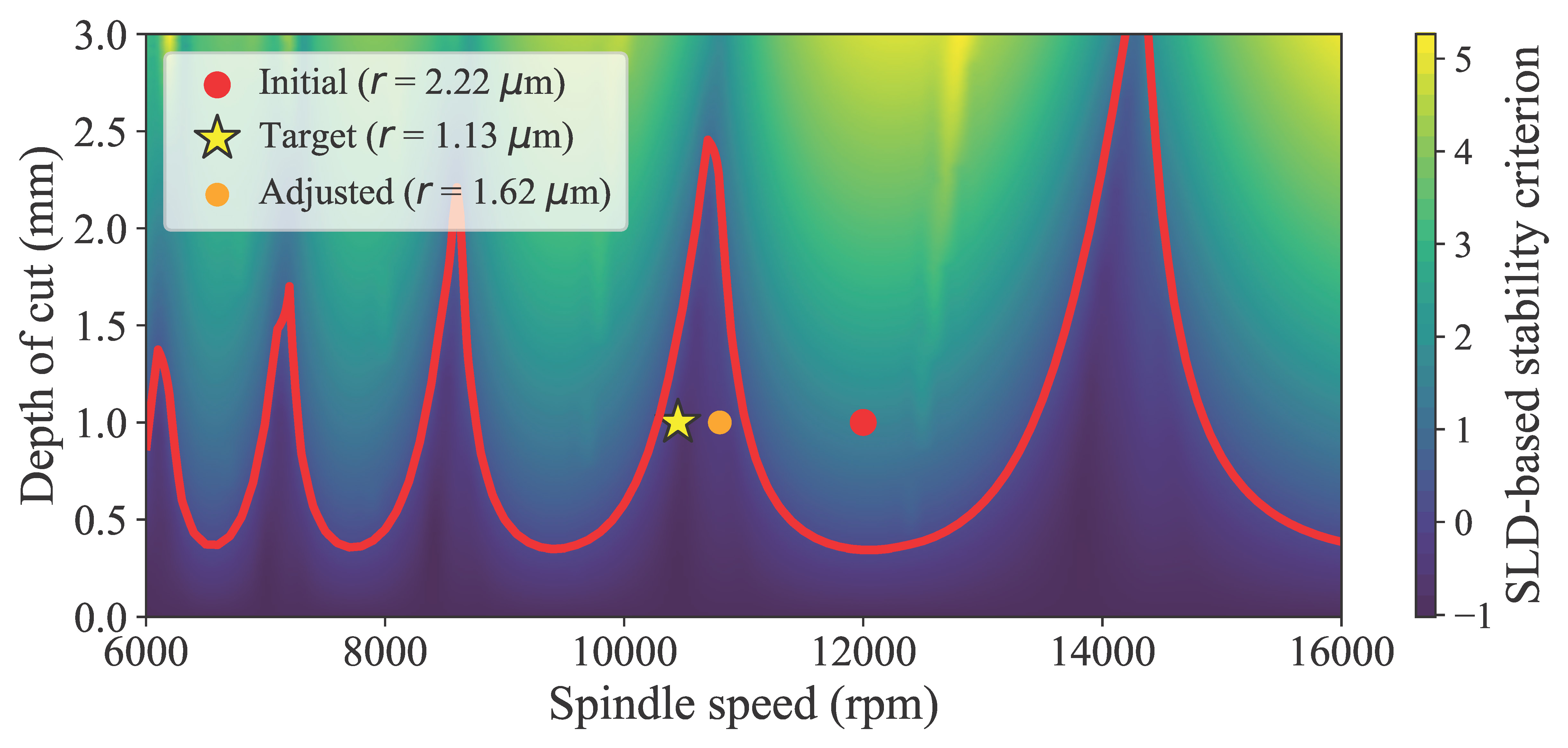}
	\vspace{-6mm}
	\caption{Adaptive spindle speed with SLD-guided stability and surface finish quality improvement.}
	\label{fig_SLD}
	\vspace{-0mm}
\end{figure}

Given the estimated SLD and surface roughness, the spindle depth is adaptively modulated in real time to suppress chatter and enhance surface finish quality. Fig.~\ref{fig_SLD} illustrates the estimated SLD. As shown in the figure, if the current milling configuration lies within an unstable region of the predicted SLD—indicating a high likelihood of chatter—a controlled adjustment of the spindle speed is applied to transition the system into a stable operating zone. To enforce SLD-based stability, we impose a hard constraint that requires the dominant singular values of the system to remain below unity, thereby ensuring a bounded dynamic response. 

As also illustrated in Fig.~\ref{fig_SLD}, there are multiple spindle speed options available in the stable regions of the SLD. More criteria are needed for optimal spindle speed selection, namely, the SLD can be used to serve as a boundary constraint for stability guarantee. In this paper, we integrate the SLD-based stability guarantee with surface roughness to simultaneously avoid chatter and enhance surface finish quality. In Fig.~\ref{fig_SLD}, the initial spindle speed is unstable, and we select the ``target'' spindle speed over the ``adjusted'' one for smaller surface roughness, $r=1.13$~$\mu$m vs. $r=1.62$~$\mu$m, though both speed options result in stable cutting. This observation highlights that multiple stable configurations within the SLD can lead to differing surface outcomes. Such flexibility is used to minimize surface roughness by selecting the optimal spindle speed from among the stable candidates.

\begin{figure*}[t!]
	\hspace{-2mm}
	\subfigure[]{%
		\includegraphics[width=0.47\linewidth]{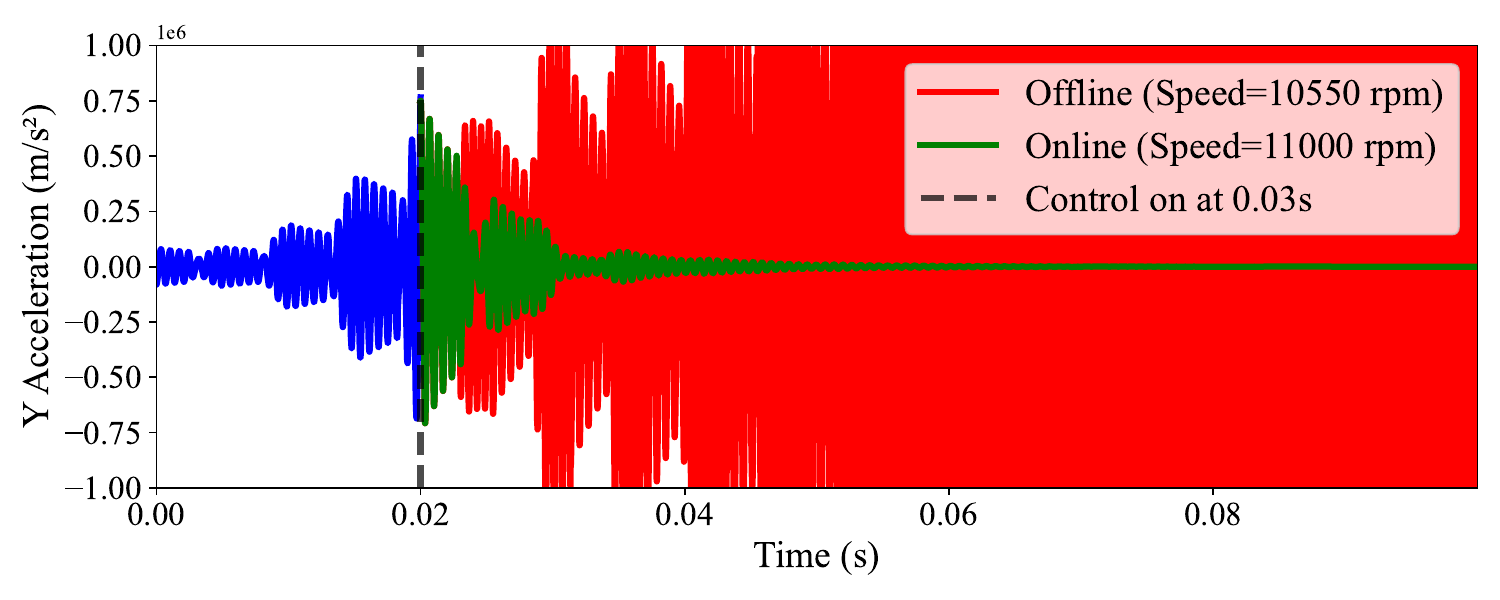}}
	\hspace{-2mm}
	\subfigure[]{%
		\includegraphics[width=0.275\linewidth]{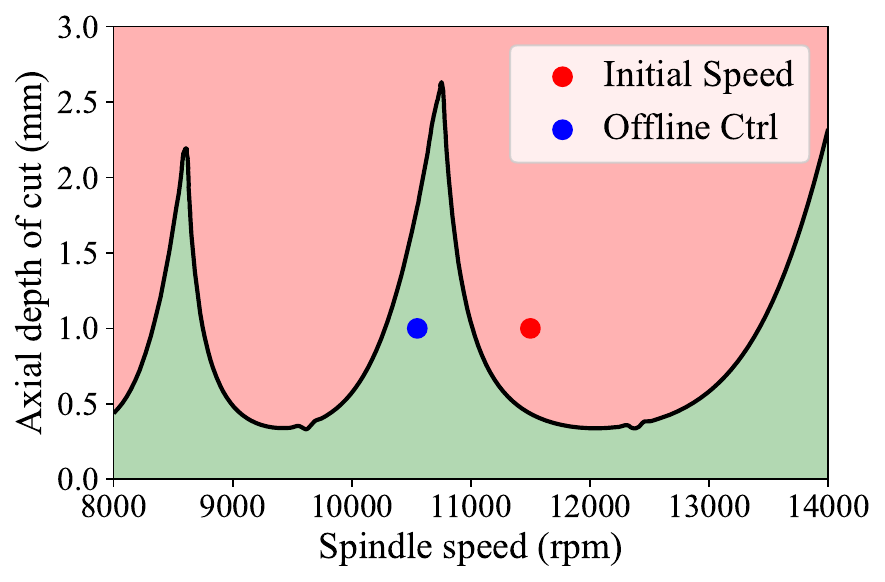}}
	\hspace{-4mm}
	\subfigure[]{%
		\includegraphics[width=0.275\linewidth]{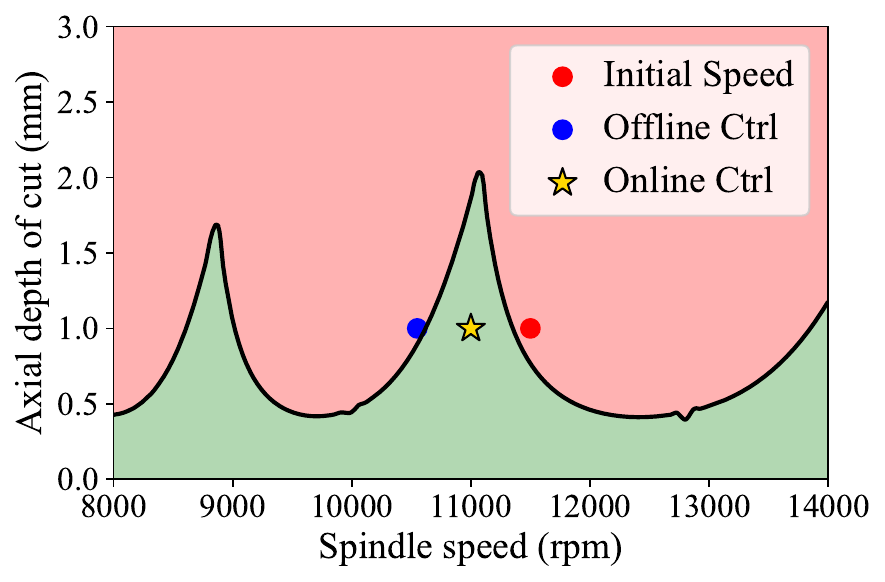}}
	\vspace{-4mm}
	\caption{Simulation results and comparison between the offline and online controls. (a) Acceleration response in milling. (b) Offline SLD. (c) Online SLD at $0.03$~s. }
	\label{fig_simulation_time}
	\vspace{-3mm}
\end{figure*}

We design the following optimization function to adapt to time-varying milling scenarios
\begin{equation}
\label{eq_ideal_spd}
\omega_{sp}^* = \arg\min\nolimits_{\omega_{sp} \in [\omega_{sp,\min}, \omega_{sp,\max}]} \mathcal{L}(\omega_{sp}),
\end{equation}
where the stability–roughness-focused loss function is defined as
\begin{align}
&\mathcal{L}(\omega_{sp}) =
\underbrace{\alpha \log \left( \lambda_{\max}(\bm{\Gamma}_i) \right)
- \beta \log \left( 1 - \lambda_{\max}(\bm{\Phi}_i) \right)}_{\mathrm{SLD-based\ stability\ criterion}} \nonumber\\
&\quad+ \underbrace{\gamma \log\left[(\omega_{sp} - \omega_{{sp}, \mathrm{current}})^2\right]} _{\mathrm{Spindle\ speed}} + \underbrace{\log(r)}_{\mathrm{Surface\ qualtiy}},
\end{align}
where $\alpha, \beta, \gamma \in \mathbb{R}_{+}$ are weighting coefficients that balance the contribution of each term. The loss function consists of the following three components:
\begin{enumerate}
\item {\em SLD-based stability enforcement}: $\alpha \log (\lambda_{\max}(\bm{\Gamma}_i))$ penalizes strong vibration response to external disturbances, while $\beta \log \left( 1 - \lambda_{\max}(\bm{\Phi}_i) \right)$ penalizes situations where the system is close to instability, discouraging operation near the stability boundary. Together, they encourage the milling status to shift within the SLD to ensure dynamic stability.
\item {\em Spindle speed jump suppression}: $\gamma \log[(\omega_{sp} - \omega_{sp,\mathrm{current}})^2]$ penalizes large changes in spindle speed, recognizing that multiple stable configurations may exist and abrupt shifts should be avoided.
\item {\em Surface roughness minimization}: $\log(r)$ promotes improved surface finish quality by minimizing the predicted roughness.
\end{enumerate}

\begin{algorithm}
\caption{Real-Time spindle speed optimization for stability and surface roughness}
\label{Algo_SLD_Ctrl}
\begin{algorithmic}[1]
\REQUIRE Estimated SLD model, surface roughness predictor, current spindle speed $\omega_{sp,\mathrm{current}}$
\ENSURE Optimized spindle speed $\omega_{sp}^*$
\STATE Acquire real-time sensor data $\bs{s}_k$ (acceleration, audio, force);
\STATE Estimate milling force $\mathbb{F}(\bs{s}_k)$ using offline dynamics model;
\STATE Extract temporal features $\mathbb{S}_n(k)$ from sensor history;
\STATE Predict surface roughness $r$ using DNN with inputs $(d_k, \omega_{sp,k}, p_k(d_k, \omega_{sp,k}), \mathbb{S}_n(k), \mathbb{F}(\bs{s}_k))$;
\STATE Evaluate SLD stability using $\lambda_{\max}(\bm{\Gamma}_i)$ and $\lambda(\bm{\Phi}_i)$;
\STATE Formulate loss function $l(v)$:
\STATE Solve: $\omega_{sp}^* = \arg\min\nolimits_{\omega \in [\omega_{sp,\min}, \omega_{sp,\max}]} \mathcal{L}(\omega_{sp})$;
\STATE Apply optimized spindle speed $\omega^*$ to the milling system;
\end{algorithmic}
\end{algorithm}

By solving the above optimization problem~\eqref{eq_ideal_spd}, we dynamically adjust the spindle speed in real time to ensure system stability while simultaneously minimizing surface roughness. The overall process is summarized in Algorithm~\ref{Algo_SLD_Ctrl}. In the following section, we will conduct both numerical analysis and experimentation to demonstrate the effectiveness of the proposed stability-enforcement and roughness suppression method.

\section{Results}
\label{sec_exp}

In this section, we present numerical and experimental results to demonstrate the performance of the SLD estimation and chatter suppression design. 

\subsection{Numerical Validations}

To evaluate the dynamic behavior of the milling system, numerical simulations were performed based on the SDM model described by~\eqref{eq_state_global} with system parameters summarized in Table~\ref{tab_sys_params}. To provide a comprehensive stability analysis, the spindle speed and axial depth of cut are varied over the ranges of $6,000–16,000$ rpm and $0–2.5$ mm, respectively. These ranges are discretized into a $200\times100$ grid to enable high-resolution evaluation of the stability boundary across a wide spectrum of cutting conditions. Fig.~\ref{fig_SLD} illustrates an SLD derived from the open-loop system described by~\eqref{eq_state_global}. The stability boundary is determined by satisfying the maximum eigenvalue of $ \bm\Phi_i(\omega_{sp}, a_p) = 1$. The stable area is defined by$ \bm\Phi_i(\omega_{sp}, a_p) < 1$, whereas the chatter occurs in regions where $\bm\Phi_i(\omega_{sp}, a_p) > 1$ is satisfied.

\begin{table}[h!]
    \centering
    \renewcommand{\arraystretch}{1.2} 
    \caption{System Parameters Used in Numerical Simulations}
    \label{tab_sys_params}
\vspace{-1mm}
    \begin{tabular}{llr}
    \hline
    {Symbol} & {Description} & {Value} \\
    \hline
    $N$ & Number of teeth & $2$ \\
    $M$ & Mass~\text{[kg]} & $0.04$ \\
    $D$ & Diameter of End Mill~\text{[mm]} & $6.35$ \\
    $a_e$ & Radial depth of cut~\text{[mm]} & $3.175$ \\
    $\zeta$ & Damping ratio & $0.011$ \\
    $\omega_n$ & Natural frequency~\text{[rad/s]} & $2\pi \times 1435$ \\
    $K_t$ & Tangential cutting coefficients [$\text{kg}/(\text{m}\cdot\text{s}^2)$] & $6 \times 10^8$ \\
    $K_r$ & Radial cutting coefficients [$\text{kg}/(\text{m}\cdot\text{s}^2)$] & $2 \times 10^8$ \\
    $a_p$ & Axial depth of cut~\text{[mm]} & Variable \\
    $\omega_{sp}$ & Spindle speed~\text{[rpm]} & Variable \\
    \hline
    \end{tabular}
\end{table}

Based on~\eqref{eq_SLD_Phi}, the estimation of SLD is given by the online prediction of system parameters with motion state data and trained model $\mathcal{N}(\cdot)$. These predicted parameters, together with the machining parameters, are then used to compute the eigenvalues and to define the SLD curve. In simulation, the offline controller was designed based on pre-identified SLD with system parameters as listed in Table~\ref{tab_sys_params}. The controller was applied to calculate the improved spindle speed according to~\eqref{eq_ideal_spd}. In contrast, the online control employed the SLD generated from the online real-time updating and prediction to adapt to system variations during the milling process. Fig.~\ref{fig_simulation_time} shows the estimated SLDs under various spindle velocities. The initial speed was $11,500$~rpm with a depth of cut $1$~mm, placing the system in the unstable region. The control actuation was activated at time $0.03$~s. Based on the offline SLD, the improved spindle speed was $10,579$~rpm, which corresponded to a stable region in the offline SLD curve shown in Fig.~\ref{fig_simulation_time}(b). However, due to the shift in the SLD curve resulting from system parameter variations at $0.03$~s, this updated spindle speed fell within the unstable region, as confirmed by the online SLD estimation in Fig.~\ref{fig_simulation_time}(c). To compensate for this shift, the online controller adjusted the spindle speed to $11,000$~rpm, thereby maintaining system stability. The acceleration response in Fig.~\ref{fig_simulation_time}(a) further confirm that the offline control failed to stabilize the system, whereas the online control successfully maintained system stability and therefore avoided potential chattering.

\subsection{Experimental Results}

Physical experiments were conducted using a benchtop CNC machine (Benchtop PRO 2424 from Avid CNC), controlled via an Acorn CNC Controller from Centroid. The sensor network comprised a 6-DOF force sensor (Mini45 model from ATI Inc.) and an accelerometer (model 8763B100 from Kistler Instrumente AG). Milling quality was assessed by measuring the surface roughness of the machined workpiece using a digital microscope (VR-3000 series from Keyence). Fig.~\ref{fig_exp_setup} shows the experimental setup. For safety considerations, the spindle speed adjustment resolution was set to 100 rpm per step. Additionally, a minimum interval of $0.2$~s was enforced between consecutive adjustments to prevent frequent adjustments and to allow sufficient settling time.

\begin{figure}[h!]
	\centering
	\includegraphics[width=0.98\linewidth]{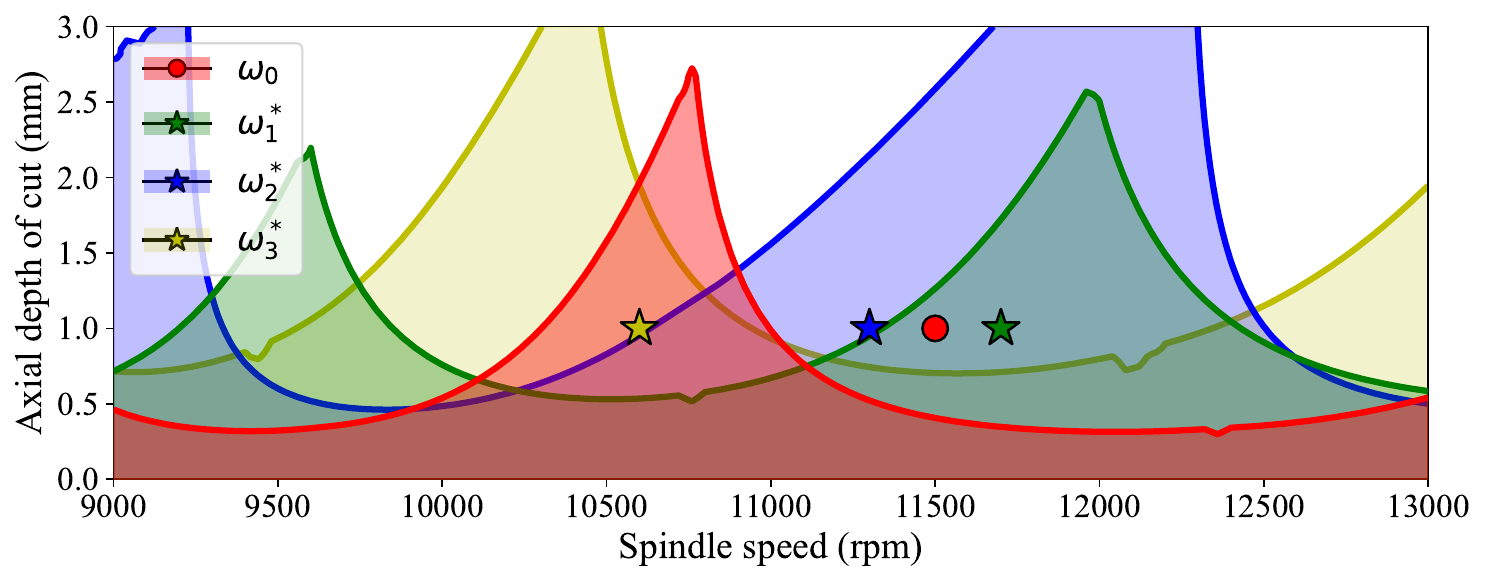}
	\vspace{-2mm}
	\caption{Online estimations of SLD and adaptive spindle speeds.}
	\label{fig_exp_sld_spd}
	\vspace{-0mm}
\end{figure}

In experiments, both offline and online controls were implemented in the CNC system to mill an aluminum block with a radial depth of cut of $3.175$~mm and feed rate $8.5$~mm/s. Fig.~\ref{fig_exp_sld_spd} shows the online estimated SLDs with the adjusted spindle speeds. With online estimation of the SLD using sensor data, the corresponding adaptive spindle speeds were addressed. Starting from an initial spindle speed of $\omega_0=11,500$~rpm, the system dynamics evolved during milling, enabling real-time SLD estimation. Based on these updates, the adaptive process control adjusted spindle speeds to $\omega_1^* = 11,700$, $\omega_2^* = 11,300$, and $\omega_3^* = 10,600$~rpm, over time. Each updated spindle speed was in the stable region of the new SLD to enforce stability and chattering avoidance.

\begin{figure}[h!]
	\centering
	\subfigure[]{%
		\includegraphics[width=1.0\linewidth]{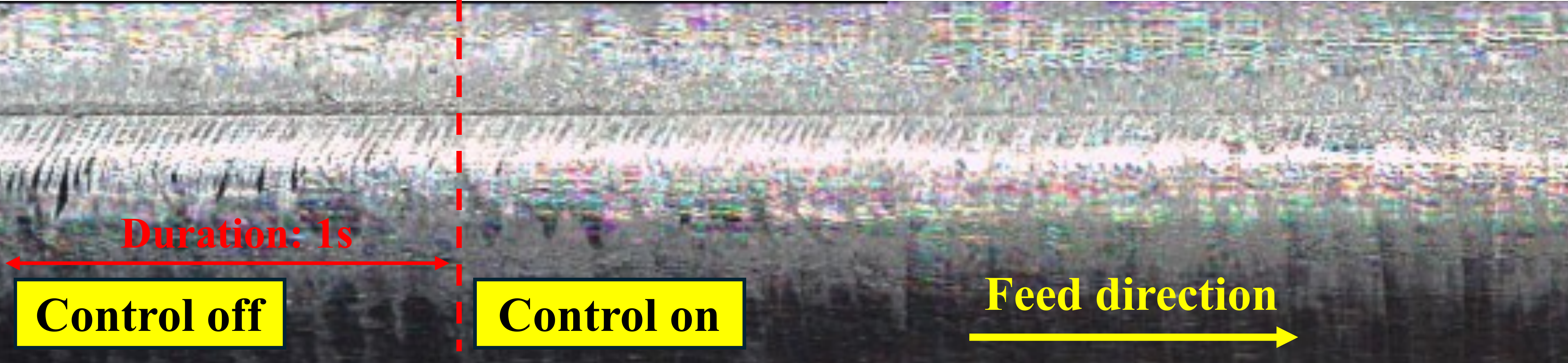}}
	\\[-0.5ex]
	\subfigure[]{%
		\includegraphics[width=0.32\linewidth]{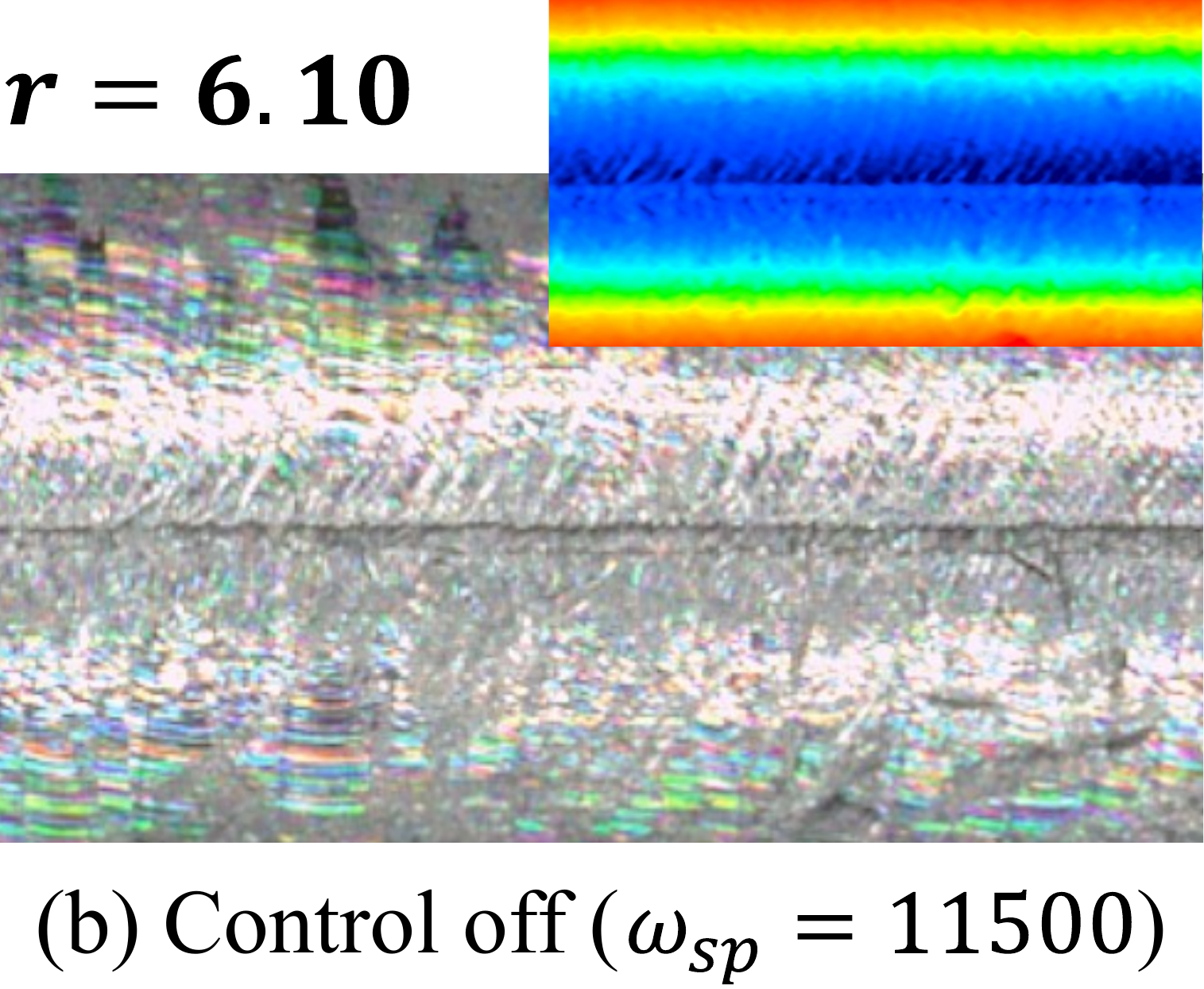}}
	\hfill
	\subfigure[]{%
		\includegraphics[width=0.32\linewidth]{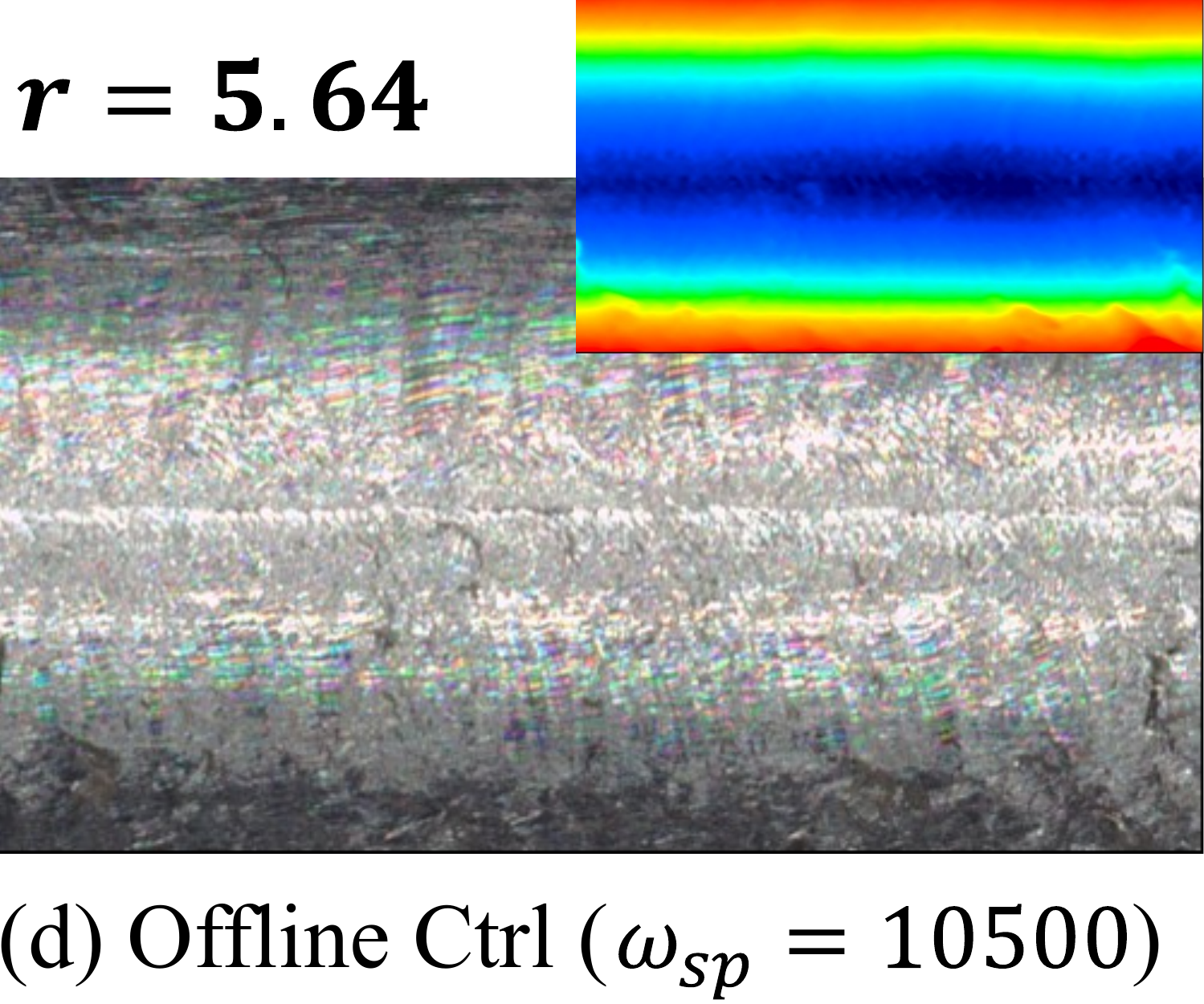}}
	\hfill
	\subfigure[]{%
		\includegraphics[width=0.32\linewidth]{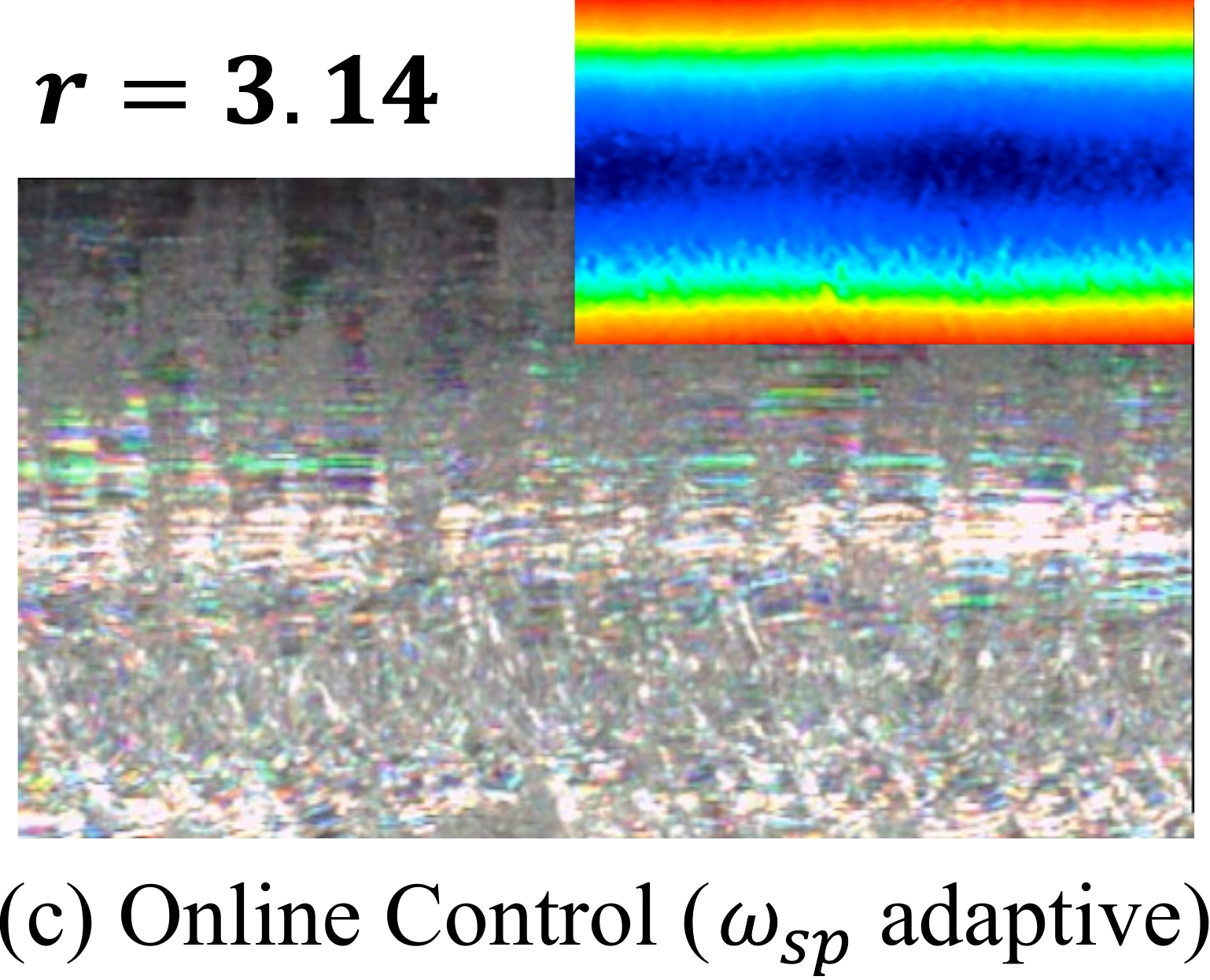}}
	
	\caption{Comparison of machined surfaces with roughness $r$ (in $\mu$m) and controllers with spindle speed $\omega_{sp}$ (in rpm). (a) Control off for the first $1$s, then stabilized by online controller. (b) Without Controller $(\omega_{sp}=11,500)$. (c) Offline controller $(\omega_{sp}=10,500)$. (d) Online controller $(\text{adaptive }\omega_{sp})$. }
	\label{fig_machined_surf}
	\vspace{-0mm}
\end{figure}

For comparison purposes, in the first $1$~s, the online control of spindle speed updating was executed, after which the online controller was activated. Fig.~\ref{fig_machined_surf}(a) shows the machined surface of a lane on the workpiece under the online adaptive spindle speed control. To quantify the control performance and surface finish quality, we used a digital microscope to measure the surface roughness $r$, with the arithmetical mean height $R_a$ employed as the roughness metric. Fig.~\ref{fig_machined_surf}(b)–(d) shows the measured $r$ under various spindle speeds with and without control. Both the offline and online controllers effectively stabilized the milling process and suppressed chatter. However, the surface finish quality under offline control was similar to that without control. The online control significantly enhanced the surface finish quality, with a $48.5\%$ roughness decrement, from $r = 6.10$~$\mu$m to $r = 3.14$~$\mu$m, highlighting the superior performance and the necessity for online SLD measurement and control updating.

\begin{figure}[h!]
  \centering
  \includegraphics[width=0.98\linewidth]{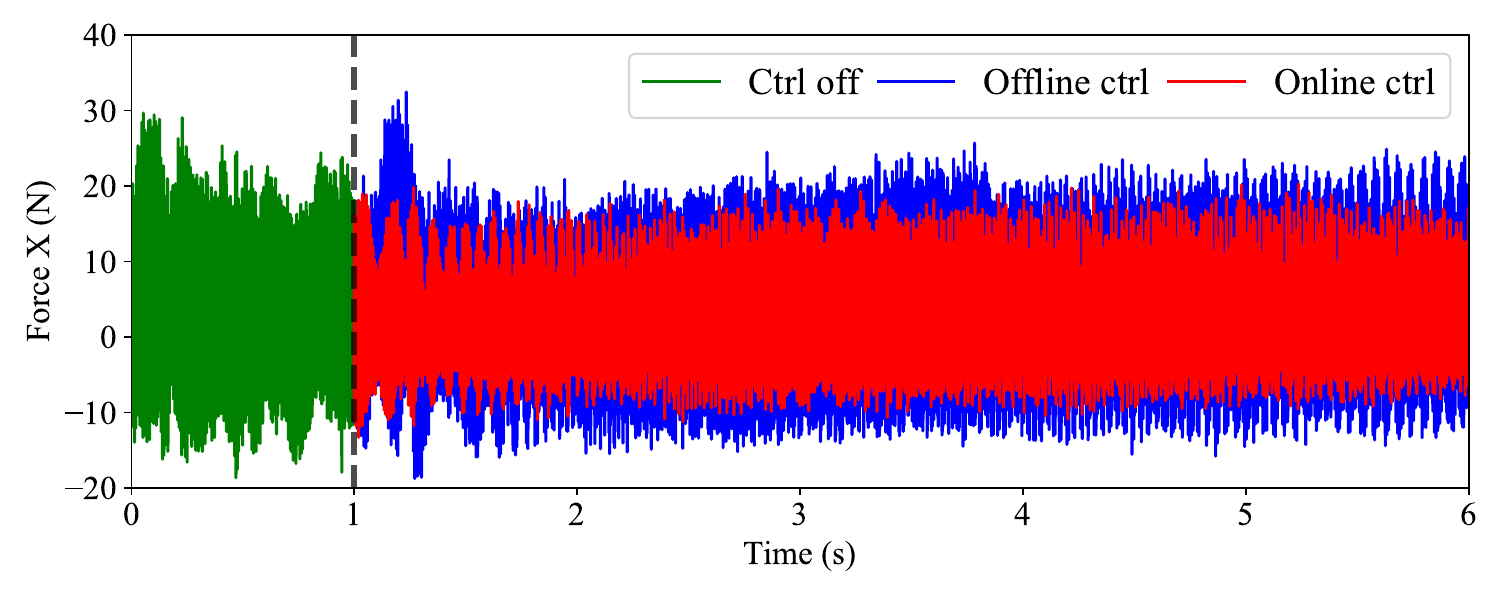}
      \vspace{-3mm}
  \caption{Cutting force measurements and comparison during the milling process under the offline and online controllers.}
  \label{fig_exp_fx}
      \vspace{-1mm}
\end{figure}

In contrast to the offline controller that adjusted the spindle speed prior using an SLD identified from constant system parameters, the adaptive online controller continuously updated the SLD using real-time sensor data. This capability resulted in superior performance, as evidenced by the improved surface quality. Fig.~\ref{fig_exp_fx} shows the measured cutting force data under both the offline and online controllers during the milling process. We also included the force measurements without active spindle speed control in the figure. From the figure, we clearly observe that the online controller effectively suppressed force growth (smaller than those under the offline controller and no control) and therefore, prevented the onset of unstable chatter. These findings highlight the enhanced effectiveness of the adaptive online controller relative to the offline approach.

\section{Conclusion}
\label{sec_Con}

This paper presented an adaptive process controller for real-time spindle speed adjustment to suppress chatter during milling. Stability analysis was conducted using the semi-discretization method, which also identified critical system parameters. A machine learning-based model was integrated to capture dynamic changes during machining by predicting system parameters from online sensor data, thereby enhancing the accuracy of SLD estimations. A data-driven chatter detection model was further developed to predict surface roughness. Finally, a real-time controller with a stability–roughness-focused loss function was proposed. Both the simulation and experimental results demonstrated its superior performance over offline adaptive process controllers. Extending the approach to diverse milling conditions and optimizing computational efficiency is among the ongoing research tasks.

\bibliographystyle{IEEEtran}
\bibliography{HuangRef}

\end{document}